\documentclass[12pt]{article}
\usepackage{epsfig,amsmath}
\usepackage{color}
\DeclareMathOperator{\mtanh}{mtanh}

\begin{document}
\newcommand{\simlt}{{\raisebox{-1.5mm}{$\stackrel{\textstyle{<}}{\sim}$}}}
\newcommand{\simgt}{{\raisebox{-1.5mm}{$\stackrel{\textstyle{>}}{\sim}$}}}
\title{Towards the construction of a model to describe the inter-ELM evolution of the pedestal on MAST}
\author{
D.\thinspace Dickinson$^{1,2}$,
S.\thinspace Saarelma$^{   2}$,
R.\thinspace Scannell$^{   2}$,\\
A.\thinspace Kirk$^{  2}$,
C.M.\thinspace Roach$^{  2}$ and 
H.R.\thinspace Wilson$^{   1}$}
\maketitle

\noindent
$^1$ York Plasma Institute, Dept of Physics, University of York, Heslington, York, YO10 5DD, UK \\
$^2$ Euratom/CCFE Fusion Association, Culham Science Centre, Abingdon, OX14 3DB, UK \\
\begin{abstract}
Pedestal profiles that span the ELM cycle have been obtained and used to test the idea that the pedestal pressure gradient in MAST is
limited by the onset of Kinetic Ballooning Modes (KBMs).  
During the inter-ELM period of a regularly type I ELM-ing discharge on MAST, 
the pressure pedestal height and width increase together while the pressure gradient increases by only 15 \% during the ELM cycle.
Stability analyses 
show that the pedestal region over which infinite-n ballooning modes are unstable also broadens 
during the ELM cycle.
To test the  relationship between  the width of the region that is unstable to $n=\infty$ ideal magnetohydrodynamic ballooning
modes and KBMs the gyrokinetic code, GS2, has been used for microstability analysis of the edge plasma region in MAST. 
The gyrokinetic simulations find that KBM modes with twisting parity are the dominant microinstabilities in the steep pedestal region, 
with a transition to tearing parity modes in the shallower pressure gradient 
core region immediately inside the pedestal top.
The region over which KBMs are unstable increases 
during the ELM cycle, and a good correlation is found 
between the region where KBMs dominate and the region that is unstable to infinite-n ideal ballooning modes. 
\end{abstract}
\newpage
\section{Introduction}

The reference scenario for ITER \cite{ref:ITER} is the high confinement mode (or H-mode) \cite{ref:HMODE}.  
The improved confinement is the result of a narrow transport barrier that forms at the plasma edge.  
Due to the strong coupling between pedestal and core confinement, which has been observed in many devices \cite{ref:RYTER2001}, 
the fusion performance of ITER will strongly depend on the achievable pedestal pressure. 
Therefore, an accurate prediction of the pedestal height is essential for the prediction and optimization of ITER performance.  
A class of instabilities called Edge Localized Modes or ELMs {\cite{ref:CONNOR}},
{\cite{ref:SUTTROP}} limits the maximum pressure pedestal height, {$P_{\rm ped}$}, that can be achieved for a given pedestal width.
Hence, understanding the width of the barrier is very important in determining the maximum pressure that can be achieved.   
An MHD stability analysis has been used to show that in order for ITER to attain a temperature pedestal of 4keV, 
which is thought to be required for ITER to attain its fusion power output objective, 
the pedestal width would have to be 2.5\% of the minor radius \cite{ref:Osborne2002}, \cite{ref:Snyder2004}.
Several empirical scalings for the pedestal width ($\Delta_{\rm ped}$) have been proposed based on
dimensionless parameters such as the normalized poloidal ion Larmor radius ($\rho^*_{\rm pol}$) or
pressure ($\beta_{\rm pol}$), but a strong co-linearity between pedestal values of $\beta_{\rm pol}$ and $\rho^*_{\rm pol}$ 
limits the power of these datasets to discriminate between very different scalings: e.g. 
early data from DIII-D could be fit equally well with either $\Delta_{\rm ped} = (\rho^*_{\rm pol})^{0.66}$ or 
$\Delta_{\rm ped} = (\beta_{\rm pol})^{0.4}$ \cite{ref:DIIID}. 
Experiments on JT-60U using hydrogen and deuterium plasmas found a
scaling of the form $\Delta_{\rm ped} \propto (\rho^*_{\rm pol})^{0.2}(\beta_{\rm pol})^{0.5}$ {\cite{ref:JT60U}}. 
Recent similarity experiments on DIII-D, ASDEX Upgrade and JET also obtain a pedestal width scaling, in normalised poloidal flux, of the form $\Delta_{\Psi_N} \propto \beta_{\rm pol}^{0.5}$ {\cite{Beurskins_POP2011}}.
Previous studies on MAST have shown  no evidence for an increase of the width with $\rho^*_{\rm pol}$ but a clear increase of width with $\beta_{\rm pol}$ \cite{kirk2009}.  
Fits of the distributions of MAST pedestal width measurements to the form $\Delta_{\rm ped} = A \beta_{\rm pol}^B$ yielded $B\approx0.5$. 
A similar scaling of the pedestal width has also been observed on DIII-D, C-MOD and AUG \cite{ref:snyderiaea}. 
The scaling $\Delta_{\rm ped} \propto \beta_{\rm pol}^{0.5}$ has been used to provide basic input to the EPED model for pedestal evolution \cite{ref:EPED1}.
The EPED model proposes that: drift wave turbulence is suppressed in the pedestal region by sheared flow;   kinetic ballooning mode (KBM) turbulence constrains the pedestal to a critical normalized
pressure gradient; and an ELM is triggered by the onset of the intermediate n ($\sim 10 - 20 $) peeling-ballooning mode instability. 
This model provides a prediction for the pressure pedestal width and height just before the ELM crash.
EPED was successfully compared to pedestal conditions in DIII-D and a range of other tokamaks \cite{ref:snyderiaea},\cite{ref:EPED1}.
The model has since been extended to exploit the fact that 
KBMs are predicted to have an onset condition similar to that of 
ideal infinite-n ballooning modes \cite{groebner}, where $n$ is the toroidal mode number.
In the more recent version of EPED, the infinite-n ideal MHD ballooning mode code, BALOO  \cite{baloo},
has been used to compute the threshold pressure gradient set by the condition that either 1\% of the pressure
profile in normalized poloidal flux \cite{groebner} or 50\% of the steep pedestal gradient region {\cite{snyder_NF2011}} is unstable to infinite-n ideal
ballooning modes.  This technique has found good agreement between the  predicted and observed pedestal widths and heights obtained prior to 
type I ELMs on DIII-D \cite{groebner} and JET \cite{Beurskens}.

In this paper, the ideas incorporated in the EPED model have been tested using data from the spherical tokamak MAST. 
In order to construct a pedestal model for MAST, the evolution of the pedestal profiles has
been studied as a function of time during the inter-ELM period, and this is described in Section 2.  In Section 3 the HELENA code 
\cite{HELENA} has been used to reconstruct equilibria during the ELM cycle and to determine infinite-n ideal 
ballooning stability, and the ELITE code \cite{ELITE,ELITE2} has been used to calculate the finite-n stability boundary.  
In Section 4 the local gyrokinetic code, GS2 \cite{KOTSCHENREUTHER_CPC1995}, has been used  
to perform linear microstability analysis in the pedestal region and
to test whether the region unstable to  
infinite-n ballooning modes is related to the region where KBMs are unstable in MAST
(i.e. to test whether ideal infinite-n MHD ballooning analysis can be used 
as a proxy for KBM stability in MAST).

\section{\label{sec:data} Inter-ELM pressure gradient evolution on MAST}

The data examined in this paper were obtained from a set of three MAST discharges \#24459, \#24452 and \#24763. All discharges were in a double null configuration with a plasma current of 850 kA, a toroidal magnetic field on axis of 0.585 T and heated by 3.4 MW of neutral beam injection. The timeslices examined were obtained during the inter-ELM period of a regularly type I ELM-ing discharge.

The data used in this paper were taken from a  
Thomson scattering system with 130 spatial points \cite{scannell2008, scannell2010}. 
This system combines eight Nd:YAG lasers and measures T$_e$ and 
n$_e$ profiles approximately every  4.2ms. This sampling rate is not normally sufficient to diagnose an individual ELM 
period (6-12ms), but arranging the profiles obtained over a number of 
plasmas as a function of time with respect to the ELM cycle allowed the general cycle of periodic ELMs to be diagnosed.
Each spatial point in the Thomson scattering system has a scattering length of 10mm. 
At the Low Field Side (LFS) edge of the plasma this results in 10mm 
radial resolution, $\Delta_R$, as the laser path is in an approximately 
radial direction. Close to the High Field Side (HFS) edge of the plasma, 
the laser is at an angle to the radial direction and this results in 
much higher (3-5mm) radial resolution. 

Measured edge profiles were fitted to an analytic parameterisation,
$P(r;\vec{a})$, convolved with the instrument function to account for the spatial resolution of the diagnostic. 
$P(r;\vec{a})$ is based on a modified hyperbolic tangent function, $\mtanh$, and is defined {\cite{scannell2011}}:
\begin{equation}
	P(r;\vec{a})  =  \frac{a_{\rm ped}-a_{\rm sol}}{2}\left[ \mtanh\left(\frac{a_{\rm etb}-r}{2a_{\Delta}},a_{\rm slope}\right) + 1 \right] +a_{\rm sol} \label{eq:fit},
\end{equation}
where 
\[ \mtanh(r,a)  =  \left( \frac{(1+a r)e^{r} - e^{-r}}{e^{r} + e^{-r}} \right), \]
using as fit parameters: pedestal height,  $a_{\rm ped}$; scrape off layer height, $a_{\rm sol}$; transport barrier position, $a_{\rm etb}$; width, $a_{\Delta}$; 
and core slope, $a_{\rm slope}$. The pedestal width for profile quantity $a$, $\Delta_{a}$, is determined from these parameters as $\Delta_{a} = 4 a_{\Delta}$.
Electron temperature and density pedestal width measurements on the LFS were typically close to the limits of spatial resolution, giving 
$\Delta_{\rm Te,ne}/\Delta_R \sim 1-2$: but on the HFS there were sufficient data points in the pedestal region to reduce the uncertainties, 
giving $\Delta_{Te,ne}/\Delta_R \sim 3-4$.
All data examined in this paper were therefore obtained from measurements at the HFS of MAST. 

\begin{figure}[htb]
\begin{center}
\epsfig{file=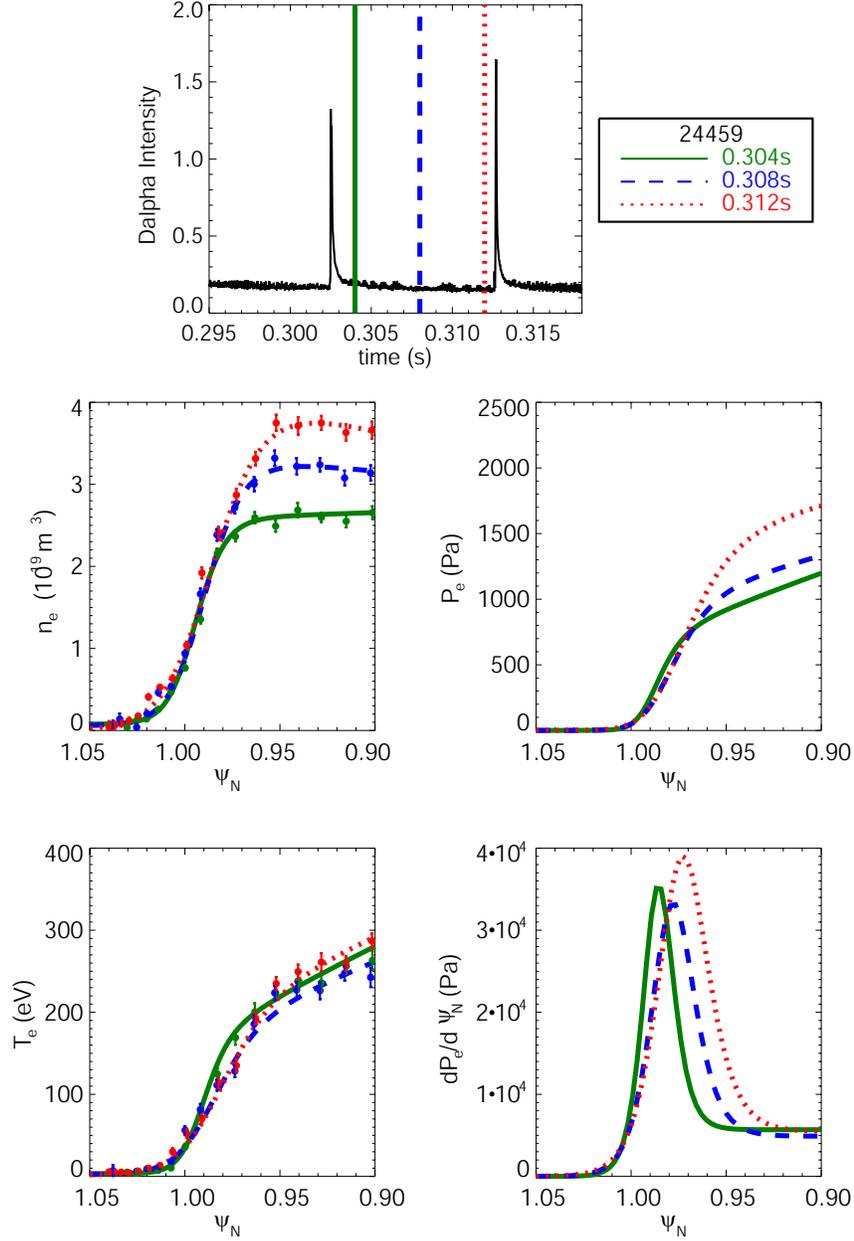,width=0.7\linewidth}
\end{center}
\caption{Profiles obtained during a single ELM cycle, showing the observed electron temperature and electron density data points in the pedestal, fitted profiles and the inferred electron pressure and electron pressure gradient profiles.}\label{fig:single_elm_cycle}
\end{figure}

Three profiles obtained during a single ELM cycle are shown in 
figure~\ref{fig:single_elm_cycle}. The T$_e$(r) and n$_e$(r) profiles 
were first fitted in real space
using the parameterisation of equation~({\ref{eq:fit}}) convolved with the instrument response function. 
These fitted profiles were then mapped to normalised poloidal flux space ($\Psi_N$), where a further fit was used to 
obtain pressure profile fit parameters in $\Psi_N$.
The electron pressure gradient, $dP_e/d\Psi_N$, was computed from this fit.

During the ELM cycle examined in figure~\ref{fig:single_elm_cycle}, 
the change in the temperature profile is small. 
The density profile shows a large change, with n$_{e,ped}$ and 
$\Delta_{ne}$ increasing throughout the ELM cycle. 
The resulting pressure pedestal change, dominated by the growing density pedestal, shows increasing $P_{e,ped}$ and expanding $\Delta_{Pe}$.
The pressure gradient profile evolution during the ELM cycle shows an expanding region of high pressure gradient (with little change in the peak pressure gradient), 
indicating that a transport barrier is expanding inwards towards the core. 

\begin{figure}[htb]
\begin{center}
\epsfig{file=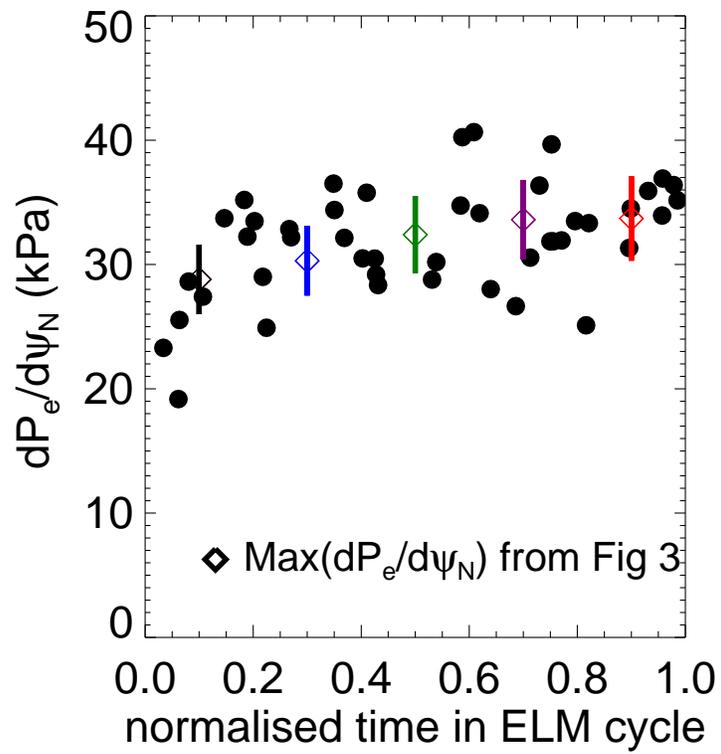,width=0.7\linewidth}
\end{center}
\caption{Evolution of the peak electron pressure gradient during the ELM cycle. The circles show the results from $\approx$50 profile measurements. The diamonds show the peak pressure gradient 
from the fitted profiles shown in figure~\ref{fig:dpdpsi_full}.
}\label{fig:dpdpsi_individual}
\end{figure}

Examining the peak pressure gradient obtained in this manner over several 
inter-ELM periods, and plotting the result as a function of time during the ELM cycle (exploiting the fact that the ELMs are periodic), results in the data shown in figure~{\ref{fig:dpdpsi_individual}}.  From this graph we see that immediately after the ELM, the drop in  pressure gradient recovers rapidly during the first 10\%  of the ELM cycle. During the remainder 
of the cycle there is only a weak upwards trend in the peak pressure gradient. 

\begin{figure}[htb]
\begin{center}
\epsfig{file=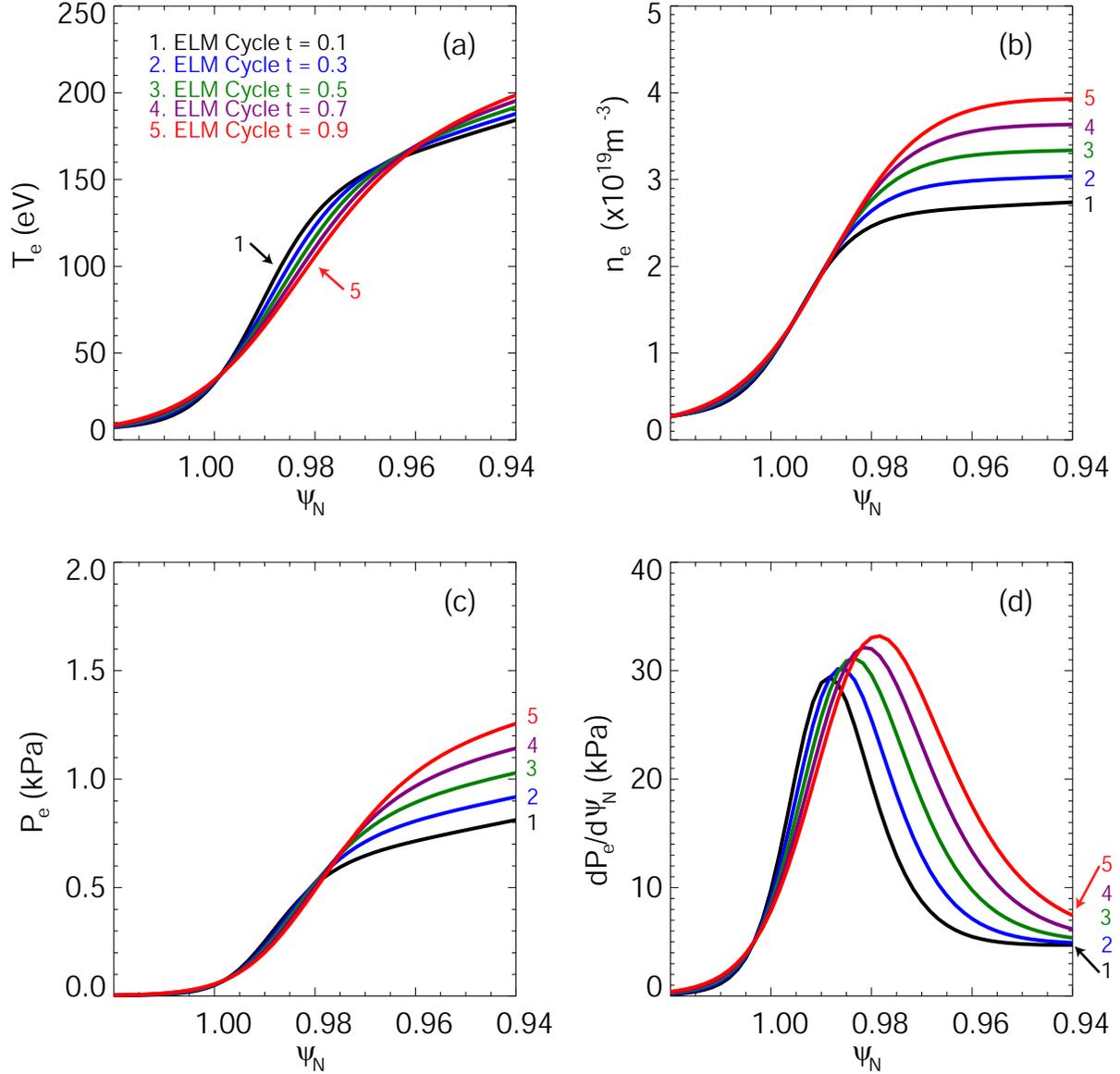}
\end{center}
\caption{Evolution over the ELM cycle of: (a) electron temperature, (b) electron density, (c) electron pressure and (d) electron pressure gradient. Profiles were obtained from 
the mtanh fit parameters evaluated at the following normalised times during the ELM cycle: $t=(0.1,0.3,0.5,0.7,0.9)$.}\label{fig:dpdpsi_full}
\end{figure}

A more complete picture of the pedestal evolution in the inter-ELM period was obtained by combining the 50 profiles from
the three MAST discharges. For each timeslice, the mtanh fit parameters for electron density, electron temperature and electron pressure profiles in $\Psi_N$ space, have been obtained as previously described. Least squares linear fits were found to represent well the time evolution of each mtanh parameter over the ELM cycle. Such linear fits were used to evaluate the mtanh profile parameters at five normalised times during the ELM cycle, $t=(0.1, 0.3, 0.5, 0.7, 0.9)$.
Table~{\ref{tab:pemtanh}} gives the interpolated values of the mtanh fit parameters for the electron pressure profile at these times, and figure~{\ref{fig:dpdpsi_full}} shows the reconstructed electron profiles. 
The peak electron pressure gradient from each of these profiles is compared with the peak electron pressure gradients determined directly from mtanh fits to individual profiles in figure~{\ref{fig:dpdpsi_individual}}. 

Figures~{\ref{fig:dpdpsi_full}}(a) and (b) show little change in the electron temperature pedestal height, 
whilst $\Delta_{Te}$, $\Delta_{ne}$ and the density pedestal height increase during the ELM cycle.
It can be seen from figure~{\ref{fig:dpdpsi_full}}(c) that the electron pressure pedestal height and width increase 
during the ELM cycle. 
Equilibrium reconstruction shows little change in the flux expansion at the HFS during the ELM cycle, and therefore 
the $\Delta_{pe}$ expansion in real space tracks that in $\Psi_N$ space.
The calculated electron pressure gradient is shown in figure~{\ref{fig:dpdpsi_full}}(d). The increase in P$_{e,ped}$ over the ELM cycle is mainly associated with 
the expansion in $\Delta_{pe}$, and only partly with a modest increase in the pressure gradient, which increases by $<20$\% over the ELM cycle.  
The high pressure gradient pedestal region expands during the ELM cycle: the innermost surface where $dp/d\Psi_N> 10$kPa moves inwards from $\Psi_N=0.975$ at $t=0.1$ to
$\Psi_N=0.95$ at $t=0.9$, while the location of the peak pressure gradient also moves inwards from $\Psi_N = 0.988$ at $t=0.1$, to $\Psi_N = 0.978$ at $t=0.9$ 
(where $t$ denotes normalised time during the ELM cycle).  
Similar observations were made during regular type I ELM-ing periods in other plasma scenarios on MAST, DIII-D {\cite{groebner}} 
and AUG {\cite{burckhart}}.

\section{\label{sec:MHD} MHD Stability Limits During the ELM Cycle}

Measured electron temperature and density profiles were used to reconstruct MHD equilibria to describe the edge plasma with high precision. 
Five such equilibria were produced to span the ELM cycle.  
Pressure profiles were obtained from the $n_e$ and $T_e$ profiles of figure~3 and model profiles for ion temperature and density, as ion measurements were not available with sufficient temporal and spatial resolution.  
A uniform effective charge state, $Z_{\rm eff} = 2$, {\footnote{$Z_{\rm eff}=2$ is consistent with typical experimental values in the MAST pedestal. The core $Z_{\rm eff}$ is typically in the range 1-1.5.}}
was assumed, for a plasma with Carbon as the main impurity species so that $n_i=n_e(7-Z_{eff})/6$.
Two ion temperature models were adopted: $T_i=T_e$ across the whole plasma including the pedestal region; and setting $T_i=T_e$ in the core, but extrapolating $T_i$ from the pedestal top to the separatrix using the electron temperature gradient from the core side of the pedestal top, to avoid a steep edge gradient in $T_i$. The plasma boundary shape, total plasma current and vacuum magnetic field were taken from MAST equilibrium reconstructions using EFIT {\cite{LAONF1985}}.
The inductively driven current profile was assumed to be fully relaxed, and the bootstrap current, which dominates at the edge, was calculated using the formula by Sauter et al. {\cite{Sauter1},\cite{Sauter2}}. These assumptions have allowed the reconstruction of high precision equilibria using the HELENA code {\cite{HELENA}}. These equilibria 
exploit the available constraints from the TS measurements self-consistently, and were used for the MHD and gyrokinetic stability analyses presented here and in Section~{\ref{sec:GK}}, respectively. 

In order to determine the limiting pressure gradients for ideal MHD instability, scans in the equilibrium pedestal pressure gradient have been performed, either by scaling the temperature pedestal height and gradient (at constant pedestal width) keeping the density profile fixed, or by scaling the density pedestal height and gradient (at constant pedestal width) keeping the temperature profile fixed. Both approaches gave similar results, but here results from the latter approach are presented, as this is closer to the experimental situation. The equilibrium scan parameter is the maximum value, $\alpha_{\rm max}$, of the normalised pressure gradient, $\alpha$, given by {\cite{miller}}:
\begin{equation} \alpha=\frac{-2\partial V/\partial \psi}{(2\pi)^2}\left(\frac{V}{2\pi^2R_0}\right)^{1/2}\mu_0 \frac{\partial p}{\partial\psi}, \end{equation} 
where $p$ is the pressure, $V$ is the volume enclosed by the flux surface, $R_0$ is the geometric centre of the plasma and $\psi$ is the poloidal flux. 
In scans the pedestal pressure gradient was varied by up to a factor 2.5 (while the equilibrium stored energy varied by $<30$\%), and the bootstrap current was always calculated self-consistently 
for each equilibrium using the appropriately modified density and temperature profiles.

Ideal MHD stability analysis was performed using two codes, depending on the toroidal mode number $n$. 
The HELENA code used for equilibrium reconstruction, was also used to determine infinite-n 
ideal ballooning stability on each flux surface {\cite{HELENA}}.
The ELITE code {\cite{ELITE,ELITE2}} was used to analyse stability of intermediate $n$ modes. In
this study the finite $n$ modes were restricted to a maximum value of $n=25$, which roughly corresponds to
the maximum $n$ observed in experimental measurements of ELM filaments in MAST \cite{kirkmastelms}.
The lowest $n$ considered was $n=5$.

\begin{figure}
\begin{center}
\epsfig{file=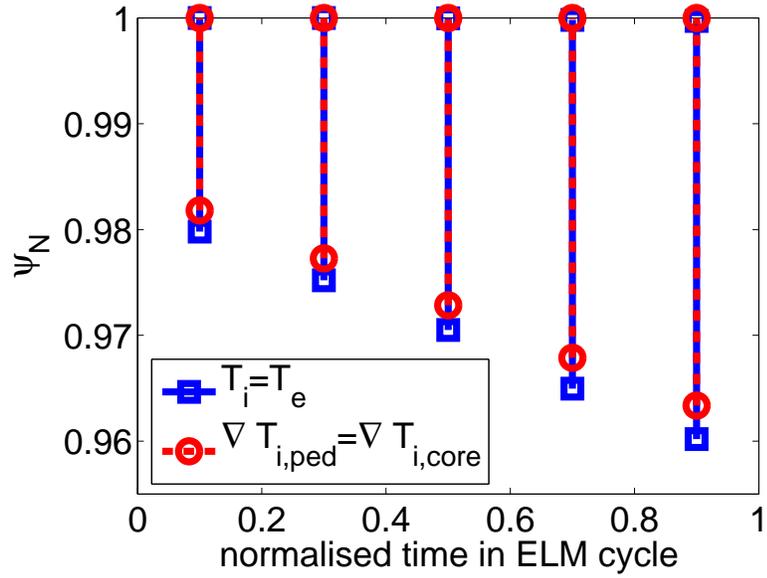,width=0.7\linewidth}
\end{center}
\caption{The unstable region for $n=\infty$ ballooning modes as a function
of the normalised relative time in the ELM cycle for assumptions $T_i=T_e$ (blue squares, solid line)
and $\nabla T_{i,{\rm ped}} = \nabla T_{e,{\rm core}}$ (red circles, dashed line).}
\label{fig:ballooningstability}
\end{figure}

All five experimental equilibria are unstable to infinite-n ideal ballooning modes. The radial region that is unstable
to $n=\infty$ modes is plotted in figure~\ref{fig:ballooningstability}
as a function of normalised time in the ELM cycle, and will be compared to the region 
unstable to KBMs in Section~{\ref{sec:GK}}.
Infinite-n ballooning stability was not found to be sensitive to the 
choice of ion temperature model in the pedestal region. All other MHD and 
gyrokinetic stability analyses were also insensitive to the ion 
temperature model, so the remaining stability results presented in this paper 
assumed $T_i =T_e$ throughout the plasma.

\begin{figure}
\begin{center}
\epsfig{file=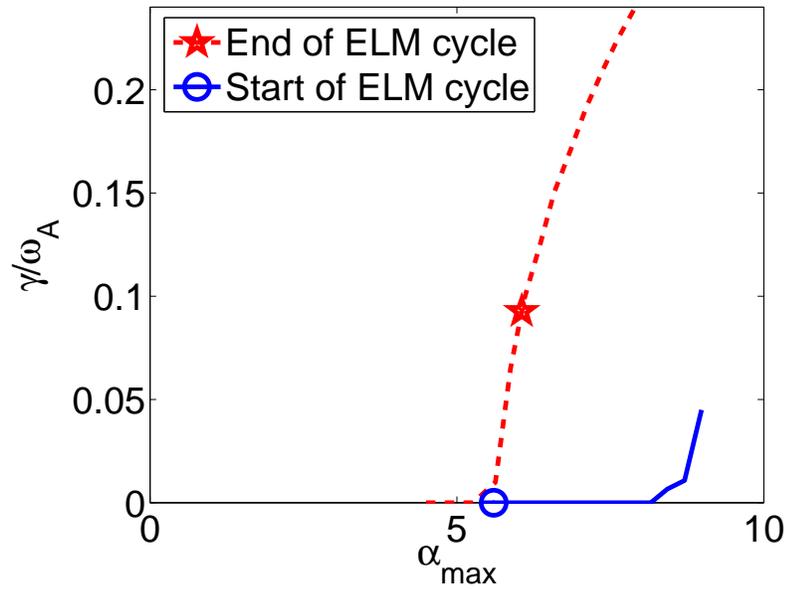,width=0.7\linewidth}
\end{center}
\caption{The growth rate of the most unstable mode, $n=25$, as a function of the maximum value of the normalised pressure gradient, $\alpha_{\rm max}$ 
for equilibria from the first (solid blue) and last (dashed red) intervals of the ELM
cycle. The circle and the star show the experimental equilibrium values of $\alpha_{\rm max}$. 
Growth rates are normalised to the Alfven frequency in the centre of the plasma ($\omega_A$).}
\label{finitenstab}
\end{figure}

Figure~{\ref{finitenstab}} shows the growth rate of the most unstable intermediate $n$
ideal MHD mode, $n=25$,  as a function of  $\alpha_{\rm max}$,
for scans around the first and last equilibria in the ELM cycle.
The threshold for the finite-n instability, $\alpha_{\rm c}$, corresponds to the $\alpha_{\rm max}$ value 
above which the mode becomes unstable with a rapidly rising growth rate. 
Figure~{\ref{finitenstab}} illustrates that while the experimental pedestal pressure gradient, $\alpha_{\rm exp}$, is far below $\alpha_{\rm c}$ at the start of the ELM cycle, 
it is slightly above $\alpha_{\rm c}$ by the end of the cycle.

\begin{figure}
\begin{center}
\epsfig{file=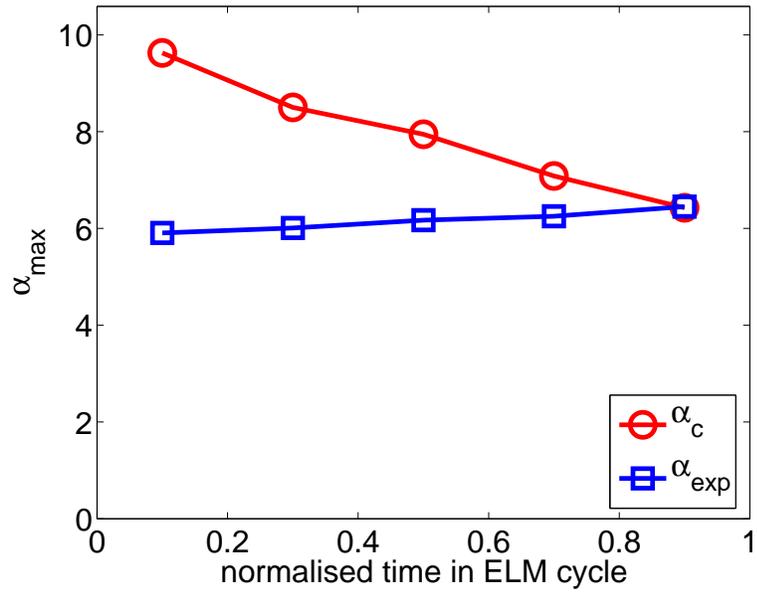,width=0.7\linewidth}
\end{center}
\caption{The maximum normalised pressure gradient in the experimental pedestal, $\alpha_{\rm exp}$, and the finite
$n$ stability limit, $\alpha_{\rm c}$, as functions of the 
normalised time through the ELM cycle.}\label{alphac}
\end{figure}

Figure~3 shows that the maximum pressure gradient
increases only slightly during the ELM cycle: the
maximum of the edge bootstrap current density ($\propto \nabla p$) also
changes very little. The main change during the ELM cycle is the broadening
of the pedestal. Figure \ref{alphac} plots $\alpha_{\rm exp}$ and $\alpha_{\rm c}$  
as functions of normalised time during the ELM cycle. While the experimental pressure gradient
changes very little, the stability limit is lowered significantly as the pedestal broadens. 
This can be understood in terms of stabilising finite $n$ corrections in ballooning theory {\cite{cht}}, 
which are larger for the equilibria with narrower pedestals from earlier in the ELM cycle.
It has frequently been found useful to present the proximity of equilibria to edge localised MHD stability boundaries using plots of 
MHD stability versus edge pressure gradient and current density ($\alpha_{max}$, $j_\phi$) {\cite{CONNOR1998}}.  Such diagrams have considerable value, but their description of stability dynamics over the full ELM cycle is limited by the difficulty of capturing the full equilibrium evolution in a simple 2-parameter model. 

\begin{figure}
\begin{center}
\epsfig{file=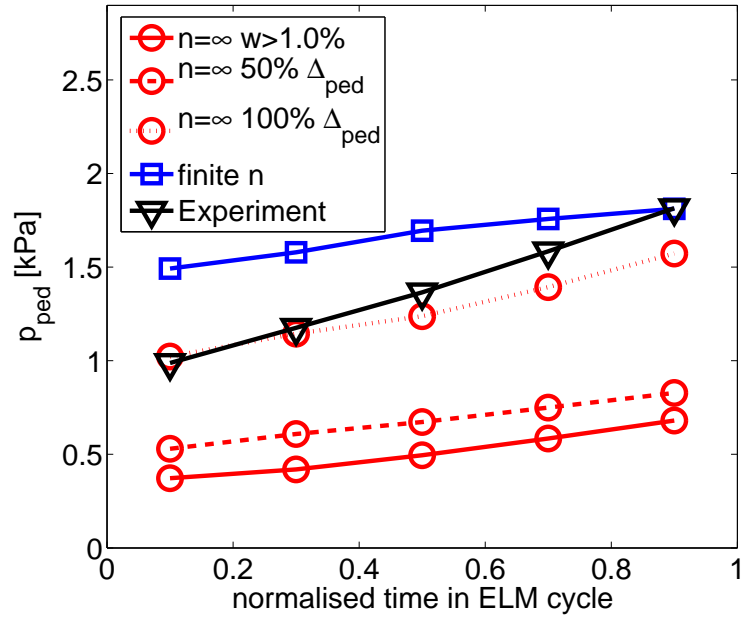,width=0.7\linewidth}
\end{center}
\caption{The pedestal top pressure of the experiment (black)
and for profiles limited by the finite $n$  
stability limit (blue), and the infinite-n ideal ballooning mode limit using three
criteria: 1\% of the total poloidal flux is unstable (solid red), 
50\% of the pedestal width is unstable (dashed red) and 100\% of the
pedestal width is unstable (dotted red).}
\label{stablimits}
\end{figure}

Figure \ref{stablimits} shows the experimental pressure pedestal height as a function of time 
during the ELM cycle. Superimposed on this plot are the stability limits derived from the
infinite-n ideal ballooning and finite-$n$ stability analyses.
The finite-$n$ stability limit is well above the experimental value at the start of the ELM cycle, 
but later it converges on the experimental value just before an ELM, which is consistent with the theory that type I
ELMs are triggered by an ideal finite-$n$ instability. 
A stability criterion, based on 100\% of the pedestal 
region\footnote{The width of the pedestal region is defined: $\Delta_{\rm ped} = (\Delta_{n_e}+\Delta_{T_e})/2$.} being unstable to infinite-n ideal MHD ballooning modes, tracks the experimental pedestal pressure height to within 20\% throughout the MAST ELM cycle.
Pedestal pressure gradient limits derived from criteria that the region unstable to $n=\infty$ ballooning modes be wider than 
(i) 1\% of the total poloidal flux, or (ii) 50\% of the pedestal, however, lie well below the experimental pedestal pressure 
throughout the cycle.

\begin{figure}
\begin{center}
\epsfig{file=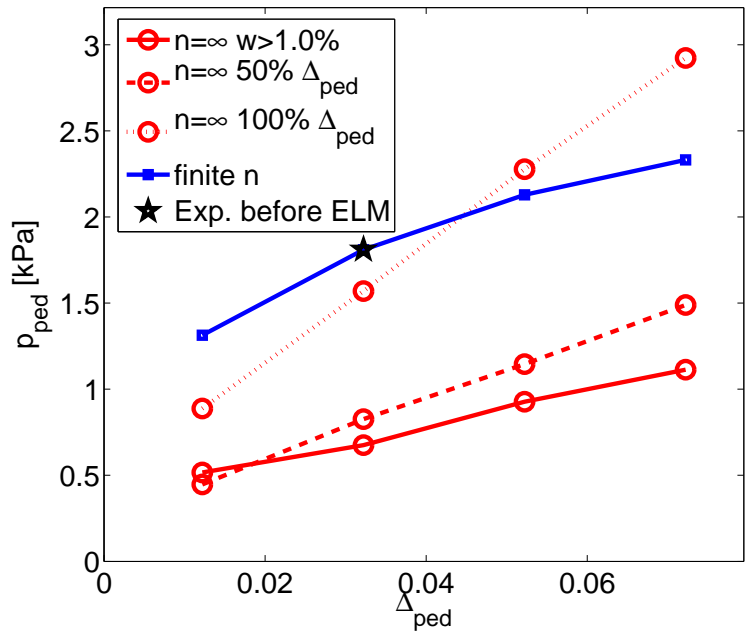,width=0.7\linewidth}
\end{center}
\caption{The pedestal top pressure as a function of the pedestal width for
profiles limited by the finite-$n$ stability 
limit (blue), and the infinite-n ideal ballooning mode
limit using three criteria: 1\% of the poloidal flux is unstable (solid red),
50\% of the pedestal width is unstable (dashed red), and 100\% of the pedestal width is unstable (dotted red). }\label{widthscan}
\end{figure}

In addition to analysing the stability of equilibria during the ELM cycle, an analysis similar to that used in the EPED model {\cite{ref:EPED1}} was repeated.
Edge stability was determined for a range of equilibria, using a scan around the reference equilibrium just prior to the ELM crash.
In this scan the density and temperature pedestal widths and the density pedestal height were scaled together around their measured values, keeping 
the density gradient constant. Then, for each pedestal width $\Delta_{\rm ped}$, the temperature pedestal height was varied to determine stability boundaries. Figure~\ref{widthscan} plots various stability limits as functions of the scan parameter $\Delta_{\rm ped}$: the finite-$n$ limit; and limits based on 1\% of the poloidal flux, and 50\% and 100\% of the pedestal, respectively, being unstable to infinite-n ideal ballooning modes. 
The pedestal top pressure limit due to finite $n$ modes increases with pedestal width, but more slowly than 
the scanned pedestal top pressure (which is $\propto \Delta_{\rm ped}$ if the pressure gradient is constant).
The limit requiring 100\% of the pedestal to be unstable to $n=\infty$ ideal ballooning modes, crosses the finite $n$ limit close to the experimental
pedestal width just before an ELM. While this could in principle provide an EPED-like model for the pedestal height and width in MAST
\footnote{A pedestal evolution model cannot be constructed for MAST using a criterion based on either 1\% of poloidal flux \cite{groebner} or 50\% of the pedestal \cite{snyder_NF2011} being unstable to infinite-n ideal ballooning modes, as the experimental pressure profile appears to be well above this limit.}, the figure of 100\% has no theoretical justification. 
It will be necessary to understand this parameter to make predictions for future devices.

\section{\label{sec:GK} Microstability analysis during the ELM cycle}
Kinetic descriptions can improve on MHD models of edge instabilities, and the gyrokinetic model is exploited here. 
The H-mode pedestal is associated with a strong pressure gradient that can 
drive electromagnetic instabilities, 
like the KBM, where the magnetic perturbations are crucial to the instability mechanism. 
Experimental measurements suggest that the KBM may play a very important role in the H-mode edge, 
by setting a hard limit on the maximum pressure gradient achievable in the pedestal region \cite{groebner}. 
Early work has shown that collisionless KBMs and ideal MHD infinite-n ballooning modes are described by related equations, 
and that these modes have similar character \cite{TANG_1}. 
This suggests that it can be reasonable to use the stability of the infinite-n ballooning mode
as a proxy for KBM stability. 
Nevertheless, finite Larmor radius (FLR) effects, only included in the kinetic treatment, 
are often stabilising \cite{TANG_2}, and the region unstable to infinite-n ideal
MHD ballooning modes can be significantly broader than the region unstable to KBMs. 
The importance of FLR effects can be measured by the parameter $\Lambda$:
\begin{equation}
\Lambda=\frac{\left(k_y \rho_i\right)^2 R}{2 L_P}
\end{equation}
where $L_P$ is the pressure gradient scale length, $k_y$ is the 
perturbation wavenumber in the flux surface and perpendicular to magnetic field lines, and $\rho_i$ is the ion Larmor radius. 
When $\Lambda \ll 1$ (corresponding to $L_P \gg (k_y \rho_i)^2 R$), FLR effects are negligible and the KBM stability limit is close to that of 
the ideal infinite-n ballooning mode: when $\Lambda \sim O(1)$, however, FLR effects become significant and should not be neglected {\cite{TANG_2}}.
This suggests that the ideal infinite-n calculation can only describe 
KBM stability consistently in regions of shallow pressure gradient, 
and raises questions as to the validity of using infinite-n ideal 
ballooning stability as a proxy for KBM stability in the pedestal region. 

In order to investigate this, a local electromagnetic gyrokinetic analysis has been used to obtain the 
microstability properties of the MAST edge plasma during the ELM cycle. 
The gyrokinetic model is based on a first order expansion in the 
small parameter $\delta= \rho_i/L$ (where $L$ is a typical equilibrium gradient scalelength),
which includes FLR corrections that are missing from ideal MHD.
Clearly gyrokinetics should be 
less accurate in the pedestal than in the core, as the pedestal 
value of $\delta$ is larger owing to shorter equilibrium gradient scale 
lengths. Using the MAST data at the time corresponding to the 
fully developed pedestal (i.e. just prior to the ELM crash), we obtain $\delta \sim 0.3$ (ie $\delta <1$) 
at $\Psi_N=0.98$, which corresponds to the surface at the mid-radius in the 
fully developed pedestal. At this location it is interesting to 
note that $\Lambda \sim 1$ for $k_y\rho_i\sim 0.2$, 
so FLR corrections should be important  for $k_y\rho_i \ge 0.2$.
While conditions in the pedestal limit the accuracy of the gyrokinetic model,
gyrokinetics is more appropriate than MHD as it includes FLR corrections.

The local gyrokinetic code, GS2, 
has been used to perform linear microstability analysis around the pedestal during the MAST ELM cycle, 
based on the data described in Section~\ref{sec:data} and the equilibria prepared for MHD stability analysis in Section~3. 
GS2 is an initial value code, 
and finds, for specified values of $k_y \rho_i$, the fastest growing microinstability 
and provides its growth rate, $\gamma$. GS2 was used to obtain the dominant growth rate spectra, 
which were captured in the range $0.07 < k_y \rho_i < 5.5$, at 12 radial locations between 
$\Psi_N=0.94$ and $\Psi_N= 0.995$, for 5 times during 
the ELM cycle, $t=(0.1, 0.3, 0.5, 0.7, 0.9)$. The characteristics of 
the dominant mode reveal the radial region where KBMs are expected to dominate,
and this can be compared directly with the region found to be unstable 
to infinite-n ideal ballooning modes using HELENA.

\begin{figure}
\setlength{\unitlength}{1.0cm}
\begin{center}
\epsfig{file=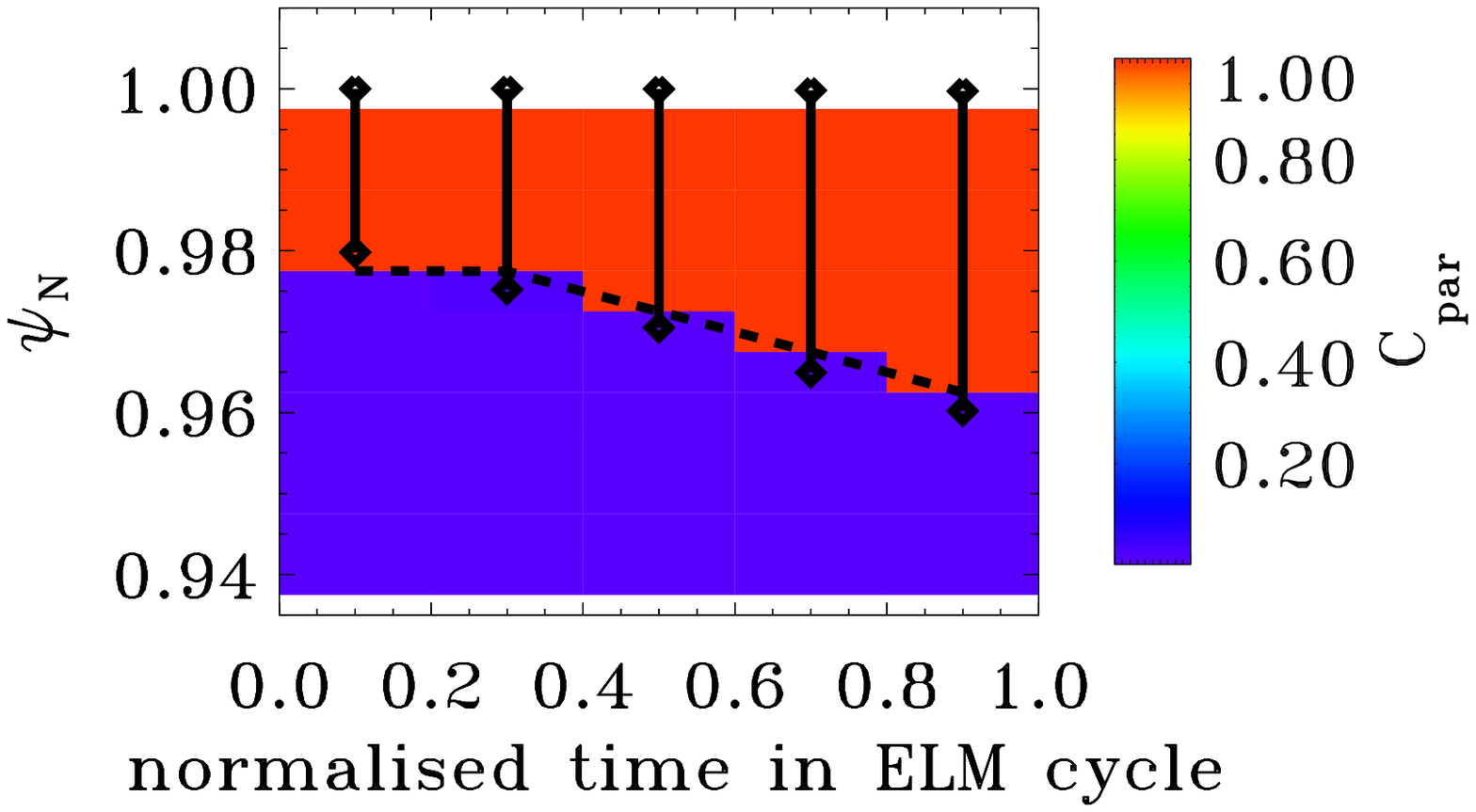,width=0.5\linewidth}
\put(-5.9,3.3){\color{black} \small \bf n=$\infty$}
\put(-4.0,3.3){\color{white} \bf \small KBM}
\put(-4.3,2.8){\color{white} \bf \small twisting}
\put(-6.,1.6){\color{white} \bf \small tearing}
\put(-8.,0.){(a)}
\put(0.,0.){(b)}
\epsfig{file=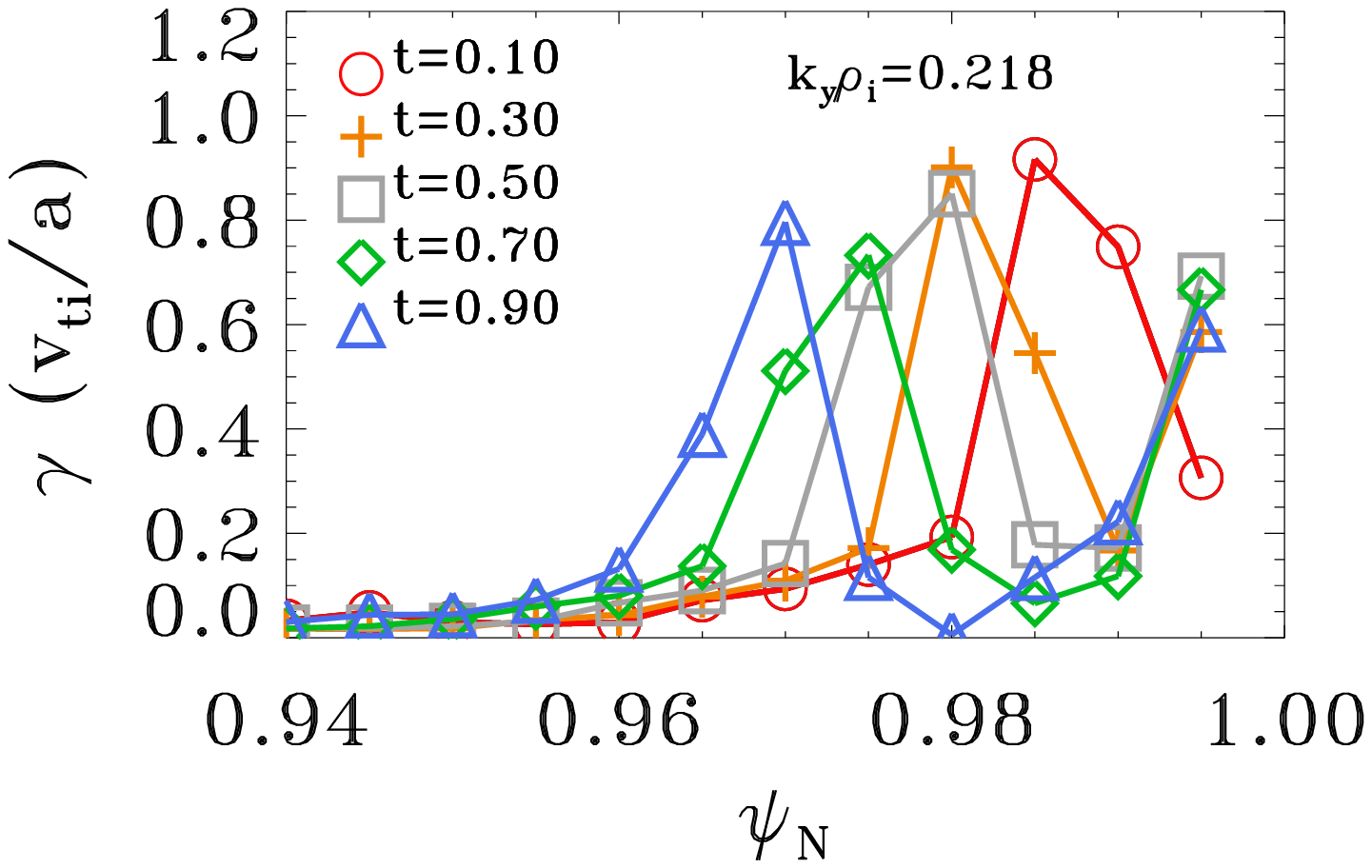,width=0.45\linewidth}
\end{center}
\caption{(a) Parity factor, $C_{\rm par}$, of the dominant mode in range $0.07 < k_y \rho_i < 5.5$, and (b) growth rate of 
the fastest growing mode at $k_y \rho_i = 0.218$, are both shown as functions of $\Psi_N$ and normalised time through the ELM cycle.
In (a) the infinite-n ideal ballooning unstable region is illustrated by black 
lines and diamonds, and the boundary between the regions where twisting
and tearing parity modes dominate is indicated by the dashed line. 
In (b) $\gamma$ is normalised to $v_{ti}/a$, 
where $v_{ti}$ is the local ion thermal velocity and $a$ is the minor radius.}
\label{fig:KBM_Contour}
\end{figure}

\begin{figure}
\setlength{\unitlength}{1cm}
\begin{center}
\epsfig{file=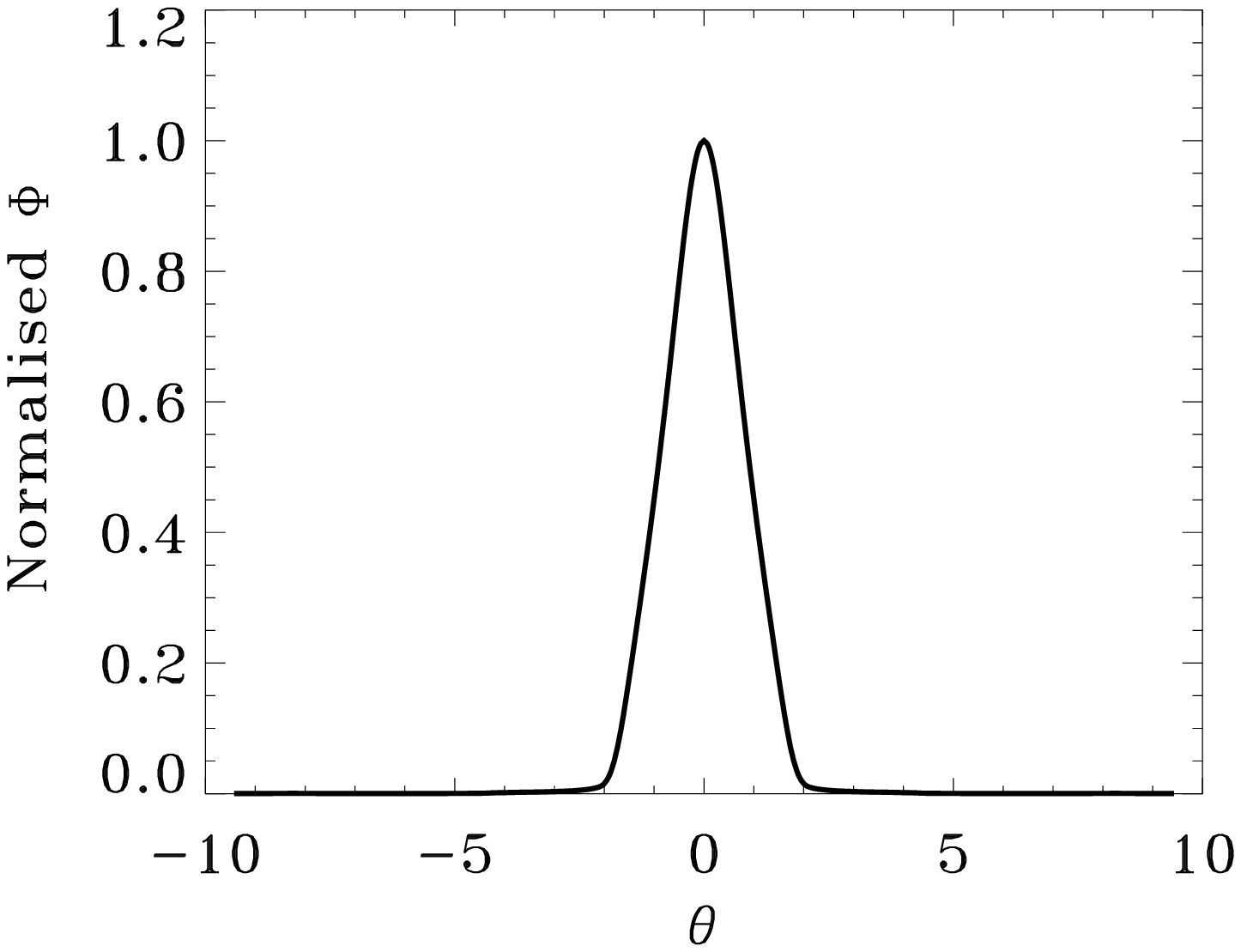,width=0.4\linewidth}
\put(-6.,0.){(a)}
\put(0.,0.){(b)}
\epsfig{file=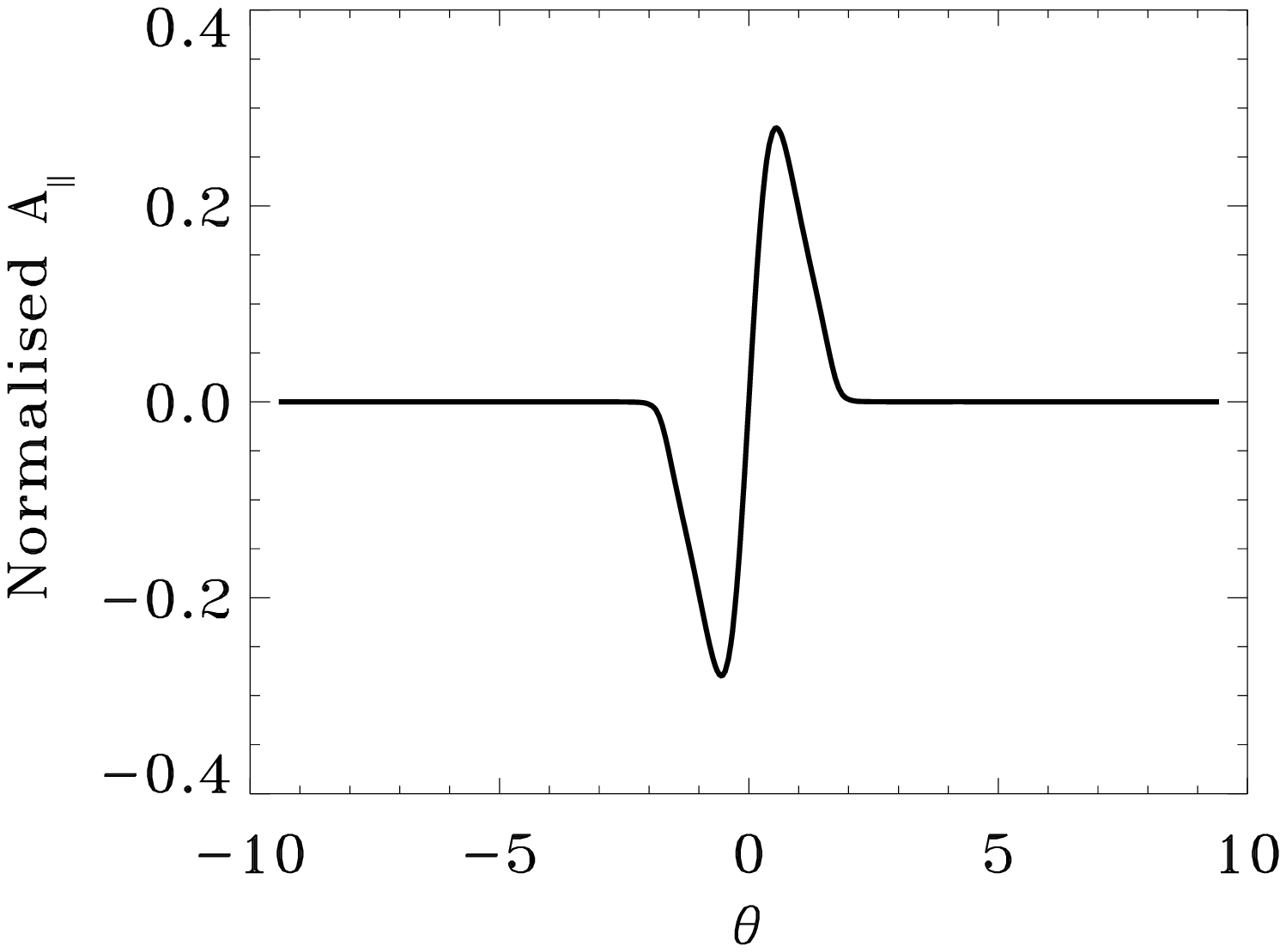,width=0.4\linewidth}
\epsfig{file=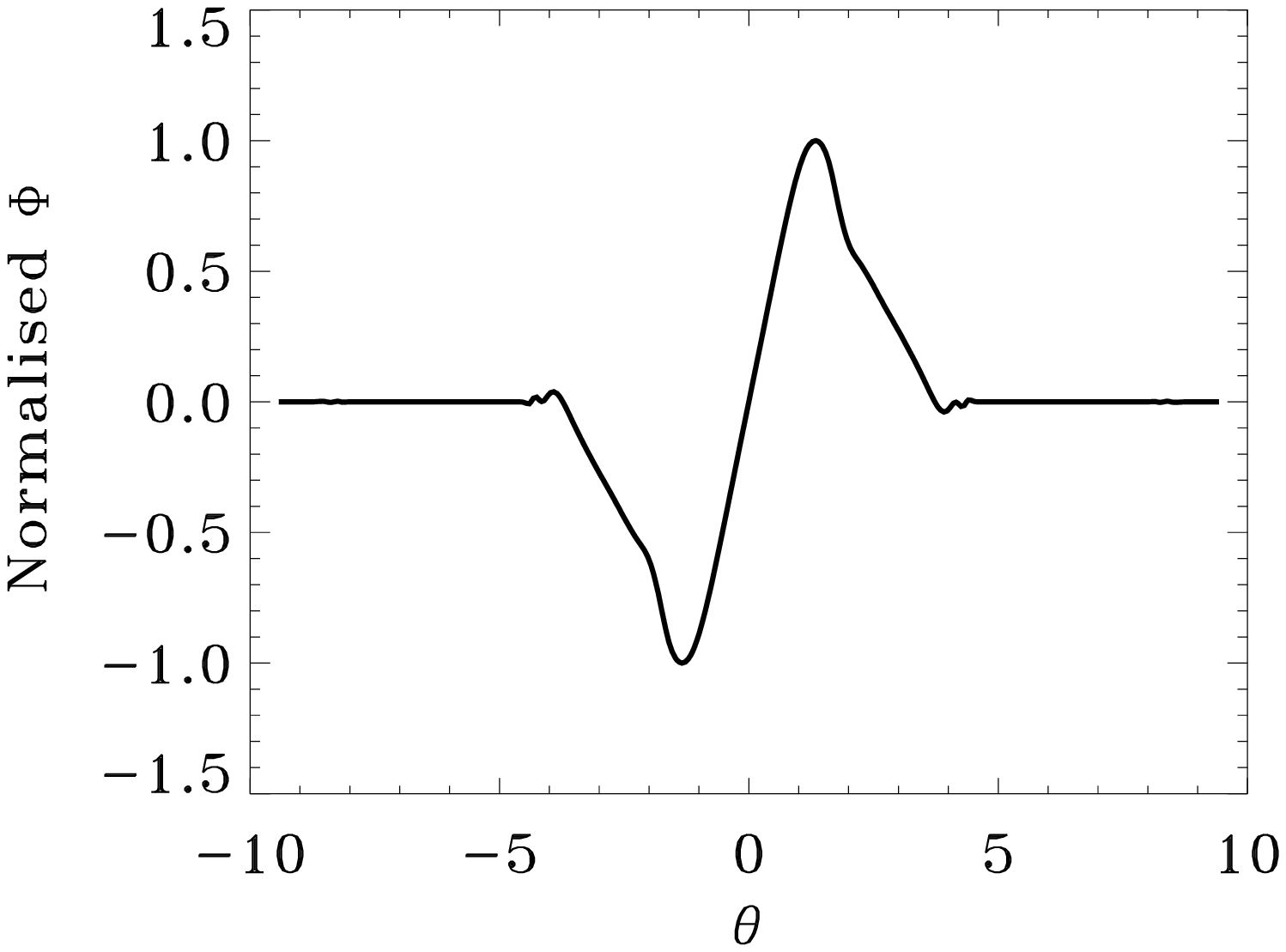,width=0.4\linewidth}
\put(-6.,0.){(c)}
\put(0.,0.){(d)}
\epsfig{file=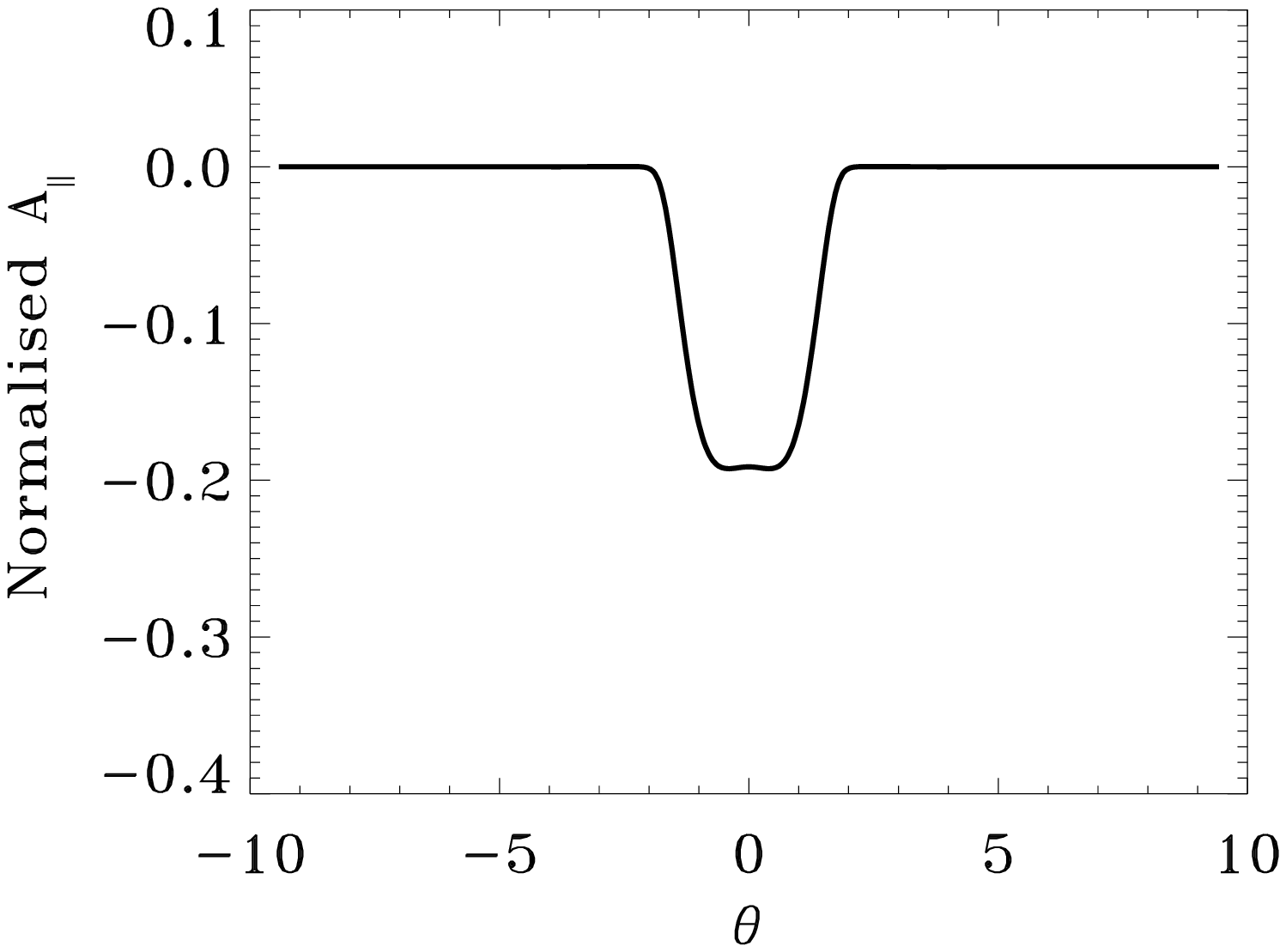,width=0.4\linewidth}

\end{center}
\caption{
Normalised real part of the $\Phi$ and $A_{\parallel}$ 
eigenfunctions for the dominant mode at the normalised ELM cycle time of $0.5$ on surfaces $\Psi_N=0.98$ and $\Psi_N=0.95$.
$\Phi$ and $A_{\parallel}$ eigenfunctions are plotted as functions of ballooning angle, $\theta$, and are normalised to 
$\Phi(\theta_{\rm max})$ and $\Phi(\theta_{\rm max})/v_{ti}$, respectively, where $|\Phi|$ peaks at $\theta=\theta_{\rm max}$. 
On the outer surface, $\Psi_N=0.98$, the dominant mode has $k_y\rho_i=0.218$, and (a) $\Phi$ and (b) $A_{\parallel}$ exhibit twisting parity.  
On the inner surface, $\Psi_N=0.95$, the dominant mode has $k_y\rho_i=3.28$, and (c) $\Phi$ and (d) $A_{\parallel}$ 
have tearing parity.}
\label{fig:Tear_KBM_Efunc}%
\end{figure}

Figure~\ref{fig:KBM_Contour}(a) shows the parity factor, $C_{\rm par}$, of the dominant mode in the $k_y$ range $0.07 < k_y \rho_i < 5.5$,
and figure~\ref{fig:KBM_Contour}(b) shows the growth rate of the fastest growing mode at $k_y \rho_i=0.218$,
which corresponds to the wavenumber of the dominant KBM instability over the ELM cycle; both quantities are illustrated as functions of radius and time 
through the ELM cycle. The parity factor is defined:
\[ C_{\rm par} = 1 - \frac{\left| \int A_{\parallel} d \theta \right|} {\int \left| A_{\parallel} \right| d \theta} \]
where $A_{\parallel}$ is the parallel magnetic vector potential.
The dominant modes are found to be KBMs with twisting parity ($C_{\rm par}=1$) in the steep pressure gradient region closer to the edge, 
with a transition to 
electron temperature gradient driven tearing parity modes ($C_{\rm par}=0$) in the shallower gradient core region inside the top of the pedestal.
The switch in mode parity is illustrated in figure~{\ref{fig:Tear_KBM_Efunc}}, which presents the dominant eigenfunctions 
of the electrostatic potential, $\Phi$, and  $A_{\parallel}$, on the $\Psi_N=0.98$ and $\Psi_N=0.95$ surfaces at $t=0.5$. 
Microtearing modes have previously been reported to be unstable at mid-radius in MAST plasmas  {\cite{APPLEGATE2004},\cite{APPLEGATE2007}}, 
close to the edge in AUG plasmas {\cite{TOLD2008}}, and in a conceptual fusion power plant based on the spherical tokamak \cite{WILSON_NF2004}. 

The infinite-n ideal ballooning unstable region (from figure~\ref{fig:ballooningstability}) is also indicated on figure~\ref{fig:KBM_Contour}(a), and this 
corresponds closely to the region where KBMs are the dominant modes. This
suggests that it is reasonable here to use infinite-n ballooning 
calculations as a proxy for the dominance of KBMs, even though $\Lambda$ is finite.
The KBM unstable region broadens and expands inwards with the strong pressure gradient, as the pedestal width increases through 
the ELM cycle.
Figure~{\ref{fig:KBM_Contour}}(b) shows that the KBM growth rate peaks in the steep gradient region close to the top of the pedestal. This peak moves inwards as the pedestal expands, and the KBM growth rate reduces at this location after the transit of the peak.

\section{Conclusions}

Pedestal profiles 
of electron temperature and electron density, obtained as a function of time during the MAST ELM cycle, have been used to test whether the pressure pedestal gradient in MAST is
limited by the onset of Kinetic Ballooning Modes (KBMs).
In this set of discharges, the electron density and electron pressure pedestal heights and widths increase during the inter-ELM period, while their gradients remain approximately constant.
Finite-n stability analyses using the ELITE code show that the finite-n stability limited normalised pressure gradient, $\alpha_{\rm c}$, lies above the experimental value, $\alpha_{\rm exp}$, initially just after an ELM, and then falls towards it just before the next ELM: i.e. rather than the experimental pressure gradient increasing towards a fixed stability boundary it is the pressure gradient stability limit that moves towards the experimental pressure gradient due to the increasing pedestal width. The region over which infinite-n modes are unstable also broadens 
during the ELM cycle. Ideal MHD stability boundary calculations
track the evolution of the pedestal height and width well, provided the limiting criterion used is that 100\% of the pedestal width be unstable to infinite-n modes.  

The gyrokinetic code, GS2, was used to test whether the MAST pedestal region unstable to KBMs corresponds closely to the region that is unstable to infinite-n ideal ballooning modes.  
KBM modes with twisting parity were found to be the dominant microinstabilities in the steep pedestal region, 
with a transition to  tearing parity modes in the shallower pressure gradient core region immediately inside the pedestal top.
The region over which KBMs dominate increases during the ELM cycle, and  closely corresponds to the region unstable to infinite-n ideal ballooning modes. 
The impact of sheared flow on 
stability has not been considered here, and an assessment of its possible impact is underway.

\bigskip
\begin{center} {\bf Acknowledgement} \end{center}

This work, part-funded by the European Communities under the contract of 
Association between EURATOM and CCFE, was carried out within the 
framework of the European Fusion Development Agreement. 
The views and opinions expressed herein do not necessarily 
reflect those of the European Commission. 
This work was also part-funded by the RCUK Energy Programme under 
grant EP/I501045. 
Gyrokinetic simulations were performed on the 
supercomputer, HECToR, which was made available through EPSRC grant EP/H002081/1.

\clearpage

\begin{table}[t]
\begin{center}
\begin{tabular}{lccccc}
$t$ & $P_{\rm e,etb}$ ($\psi_N$) & $100 \; P_{{\rm e},\Delta}$ ($\psi_N$)  & $P_{\rm e,ped}$ (Pa) & $P_{\rm e,slope}$ (Pa) & $P_{\rm e,sol}$ (Pa) \\
0.1 & 0.988$\pm$0.0009 & $0.527 \pm 0.027$ & 557.5$\pm$40.9 & 0.167$\pm$0.010 & 0.56$\pm$0.31 \\
0.3 & 0.986$\pm$0.0008 & $0.606 \pm 0.026$ & 682.5$\pm$39.1 & 0.166$\pm$0.009 & 0.47$\pm$0.29 \\
0.5 & 0.983$\pm$0.0009 & $0.686 \pm 0.029$ & 807.6$\pm$43.4 & 0.164$\pm$0.011 & 0.38$\pm$0.30 \\
0.7 & 0.981$\pm$0.0011 & $0.766 \pm 0.035$ & 932.7$\pm$52.3 & 0.162$\pm$0.013 & 0.29$\pm$0.34 \\
0.9 & 0.979$\pm$0.0014 & $0.846 \pm 0.042$ & 1058.$\pm$63.9 & 0.160$\pm$0.016 & 0.20$\pm$0.40 \\
\end{tabular}
\caption{The five mtanh fit parameters for electron pressure at $t=(0.1, 0.3, 0.5, 0.7, 0.9)$, with 
uncertainties from the least squares linear fits to the evolution of these parameters during the inter-ELM period. \label{tab:pemtanh}}
\end{center}
\end{table}

\clearpage

\clearpage

\clearpage

\clearpage

\clearpage
\end{document}